\documentclass[fleqn,usenatbib]{mnras}
\usepackage{mathptmx}
\usepackage[varvw]{newtxmath}  
\usepackage[T1]{fontenc}
\usepackage{ae,aecompl}
\usepackage{times}
\usepackage{amsmath}
\usepackage{upgreek}
\usepackage{xcolor}
\usepackage{graphicx}
\usepackage{pdflscape}
\usepackage{multicol} 
\usepackage{amsmath,bm}
\usepackage{caption}

\begin{document}

\title[The significant impact of tidal effects from neighboring material on the evolution of molecular clouds]{An examination of large-scale galactic effects on molecular cloud properties in NGC 628 : The significant impact of tidal effects from neighboring material on the evolution of molecular clouds}

\author[J. W. Zhou]{
J. W. Zhou \thanks{E-mail: jwzhou@mpifr-bonn.mpg.de}$^{1}$
Sami Dib $^{2}$
\\
$^{1}$Max-Planck-Institut f\"{u}r Radioastronomie, Auf dem H\"{u}gel 69, 53121 Bonn, Germany\\
$^{2}$Max Planck Institute f\"{u}r Astronomie, K\"{o}nigstuhl 17, 69117 Heidelberg, Germany
}

\date{Accepted XXX. Received YYY; in original form ZZZ}

\pubyear{2024}
\maketitle

\begin{abstract}
The physical factors that influence the development of molecular cloud's density contrast are connected to those that affect star formation in the galaxy. For NGC 628 (M74), the proportion of high- and low-density contrast clouds initially increases with the distance to the galactic center ($R_{G}$) and then keeps relatively stable. 
Spiral arms, bubbles and magnetic fields are not responsible for the variations in density contrast observed among molecular clouds. 
The effects of shear and tides calculated from the galactic rotation curve consistently decrease as $R_{G}$ increases, and the shear effect can be neglected. 
We further studied the tidal effects of the neighboring material on each cloud using the tidal tensor analysis and the pixel-by-pixel computation, after combining molecular gas, atomic gas and stellar mass surface density maps. When $R_{\rm G} <$ 4 kpc, the tidal strengths derived from the pixel-by-pixel computation decrease as $R_{\rm G}$ increases, and then remains relatively constant when $R_{\rm G} >$ 4 kpc. This aligns well with the dependence of the proportion of high- and low-density contrast clouds on $R_{\rm G}$. Therefore, the tidal effects of neighboring material have a significant impact on the development of molecular cloud's density contrast.
A key factor contributing to the low star formation rate in the galactic center is the excessive tidal influences from neighboring material on molecular clouds, which hinder the gravitational collapse within these clouds, resulting in low density contrasts. The tidal effects from neighboring material may also be a significant contributing factor to the slowing down of a pure free-fall gravitational collapse for gas structures on galaxy-cloud scales revealed in our previous works by velocity gradient measurements.
\end{abstract}


\maketitle 

\section{Introduction}

Giant molecular clouds (GMCs) are widely recognized as the primary gas reservoirs that fuel star formation and serve as the birthplaces of nearly all stars. Gaining insight into the characteristics of GMCs is therefore crucial for deciphering the relationship between gas and star formation in galaxies \citep{Kennicutt2012-50,Schinnerer2024-62}. Observations show that the physical characteristics of GMCs systematically change depending on their location within a galaxy. This implies a connection between GMCs and their galactic environment, which influences their formation, structure, and evolution \citep{Hughes2013-779,Colombo2014-784,Miville2017-834,Sun2022-164,Kim2022-516,Pessa2022-663,Zhou2024-534}. The range of potential physical mechanisms that influence the properties of GMCs is extensive and diverse
\citep{Jeffreson2020-498,Chevance2023-534} including galactic shear and tides \citep{Ballesteros2009-395,Dib2012-758,Thilliez2014-31,Aouad2020-496,Ramirez2022-515}, interactions with spiral arms \citep{Meidt2013-779,Querejeta2024-687}, epicyclic motions driven by
the galactic rotation \citep{Meidt2020-892,Utreras2020-892,Liu2021-505}, accretion flows from galactic scales
down to GMC scales \citep{Klessen2010-520,Zhou2024PASA}, collisions between clouds \citep{Tasker2009-700,Fukui2021-73,Sano2021-73,Zhou2023-519}, stellar feedback processes \citep{Dib2009-398,Krumholz2014-243K,Barnes2022-662,Chevance2022-509} and magnetic fields \citep{Li2011-479,Crutcher2012-50,Ibanez2022-925, Ganguly2023-525,Rawat2024-528}.
These intricate interactions result in significant, measurable correlations between the characteristics of molecular clouds and both the local and global properties of their host galaxy. Investigating these correlations between clouds and their environments presents an opportunity to deepen our understanding of the physics that drive molecular cloud evolution, and, by extension, star formation and galaxy evolution \citep{Sun2022-164,Schinnerer2024-62}.

In \citet{Zhou2024-534}, we decomposed the molecular gas in the spiral galaxy NGC 628 (M47) into multi-scale hub-filament structures using the CO (2-1) line map. All molecular clouds as potential hubs were classified into three categories: leaf-HFs-A, leaf-HFs-B, and leaf-HFs-C. For leaf-HFs-C, the density contrast between the hub and the surrounding diffuse gas is not significant. Both leaf-HFs-A and leaf-HFs-B feature distinct central hubs, but leaf-HFs-A exhibit the best-defined hub-filament morphology. Leaf-HFs-A also have the highest density contrast, the largest mass and the lowest virial ratio. Leaf-HFs-C are not necessarily low-density structures, but their density is more uniformly distributed. Currently, leaf-HFs-C lack a clear gravitational collapse process that would produce a pronounced density contrast. We found a clear correlation between density contrast and the virial ratio, with the density contrast decreasing as the virial ratio increases. The density contrast ($C$) effectively gauges the degree of gravitational collapse and the depth of the gravitational potential well that shape the hub-filament morphology. In terms of understanding the development and evolutionary stage of molecular clouds, the density contrast is more indicative than density itself.

The study of the formation and evolution of three types of structures mentioned above is important to understand what are the main mechanisms that regulate star formation in galaxies.
The interstellar medium (ISM) in galaxies ranges from diffuse gas to regions of localized clumping, followed by a further increase in density due to gravitational collapse and accretion, which eventually results in star formation. The physical factors that influence the development of molecular cloud's density contrast are the same factors that affect star formation in galaxies. As shown in \citet{Zhou2024-534}, as long as $C$ is different, other physical quantities will also differ significantly. $C$ can be used to represent the star formation capability of molecular clouds. The correlation between $C$ and various physical quantities can reflect the impact of those physical processes on the star formation in molecular clouds. In this work, we are not concerned with the formation of molecular clouds, but rather with the physical processes that create variations in density contrast among molecular clouds in NGC 628. Particularly, we focus on the role of different types of tides.


\section{Data}

In this work, we use the integrated intensity map derived from the combined 12m+7m+TP CO (2$-$1) data cube of NGC 628, which is the data product in the PHANGS-ALMA survey \citep{Leroy2021-257,Leroy2021-255}. We also use 
the integrated intensity map of HI from the HI Nearby Galaxy Survey (THINGS) \citep{Walter2008-136}, and the \textit{Spitzer} IRAC $3.6$~$\mu$m image in the \textit{Spitzer} Infrared Nearby Galaxies Survey (SINGS) \citep{Dale2009-693}.

\section{Results}

\subsection{Density contrast distribution}\label{C}

\begin{figure*}
\centering
\includegraphics[width = 1\textwidth]{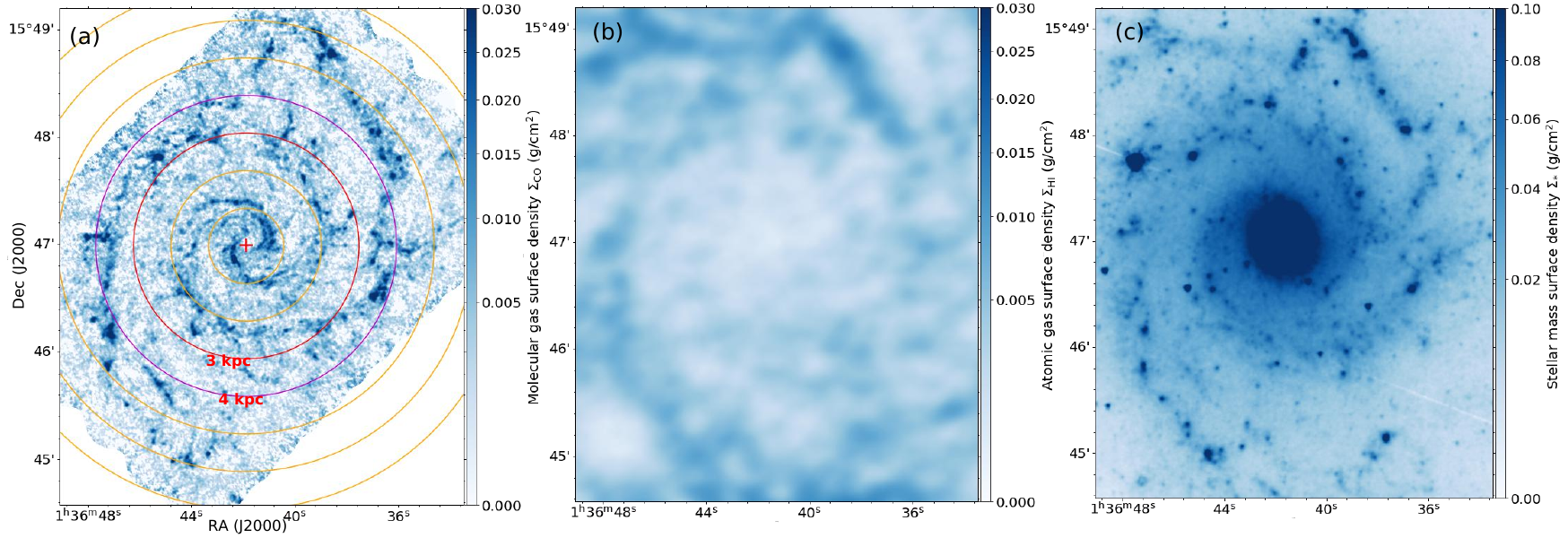}
\caption{Surface density maps of CO (2-1), HI and stellar mass.}
\label{surface}
\end{figure*}

\begin{table}
\centering
\caption{The ratio of high-density contrast clouds to low-density contrast clouds at different distances from the galactic center.}
\label{tab1}
\begin{tabular}{ccccc}
\hline
distance (kpc)	&	$C$ > 1.63	&	$C$ < 1.63	&	proportion	\\
$R_{\rm G}$ < 1	&	14	&	39	&	0.358	\\
1 < $R_{\rm G}$ < 2	&	45	&	63	&	0.714	\\
2 < $R_{\rm G}$ < 3	&	68	&	77	&	0.883	\\
3 < $R_{\rm G}$ < 4	&	94	&	68	&	1.382	\\
4 < $R_{\rm G}$ < 5	&	88	&	75	&	1.173	\\
5 < $R_{\rm G}$ < 6	&	57	&	42	&	1.357	\\
6 < $R_{\rm G}$ < 7	&	20	&	18	&	1.111	\\
\hline
\label{prop}
\end{tabular}
\end{table}

In Fig.\ref{contrast}, we can see a significant proportion of low-density contrast clouds concentrated in the galaxy center.
For all clouds, the median value of the density contrast $C$ is $\sim$ 1.63.
Based on the distance to the galaxy center ($R_{\rm G}$), we divided the galaxy into different rings. 
As shown in Fig.\ref{surface}(a), only rings with a radius of $\leq$ 4 kpc are complete. In Tab.\ref{prop}, for $R_{\rm G} \leq$ 4 kpc, the number of high-density contrast clouds rises notably as $R_{\rm G}$ increases.
In order to extend the comparison to a wider spatial range within the galaxy, we calculate the proportion of clouds with $C >$ 1.63 and $C <$ 1.63 within each ring. As shown in Sec.\ref{sbm} and Sec.\ref{Sshear}, the distributions of high- and low-density contrast clouds are relatively uniform within each ring, meaning that they are not biased toward any specific part of the ring. Despite the incompleteness of the outer rings, calculating the ratio of high- to low-density contrast clouds remains feasible.

In Tab.\ref{prop}, $R_{\rm G} <$ 3 kpc has significantly lower proportion of clouds with $C >$ 1.63, consistent with the fact that star formation is suppressed in the center of the galaxy \citep{Kruijssen2014-440}. 
When $R_{\rm G} >$ 3 kpc, the proportion is relatively stable. 
When $R_{\rm G} <$ 4 kpc, there is a clear trend where the proportion of clouds with $C >$ 1.63 increases significantly with increasing $R_{\rm G}$. Next, we will explore in detail the reasons behind the dependence of the cloud's density contrast on $R_{\rm G}$.

\subsection{Spiral arm, bubble and magnetic field}\label{sbm}

\begin{figure*}
\centering
\includegraphics[width = 0.8\textwidth]{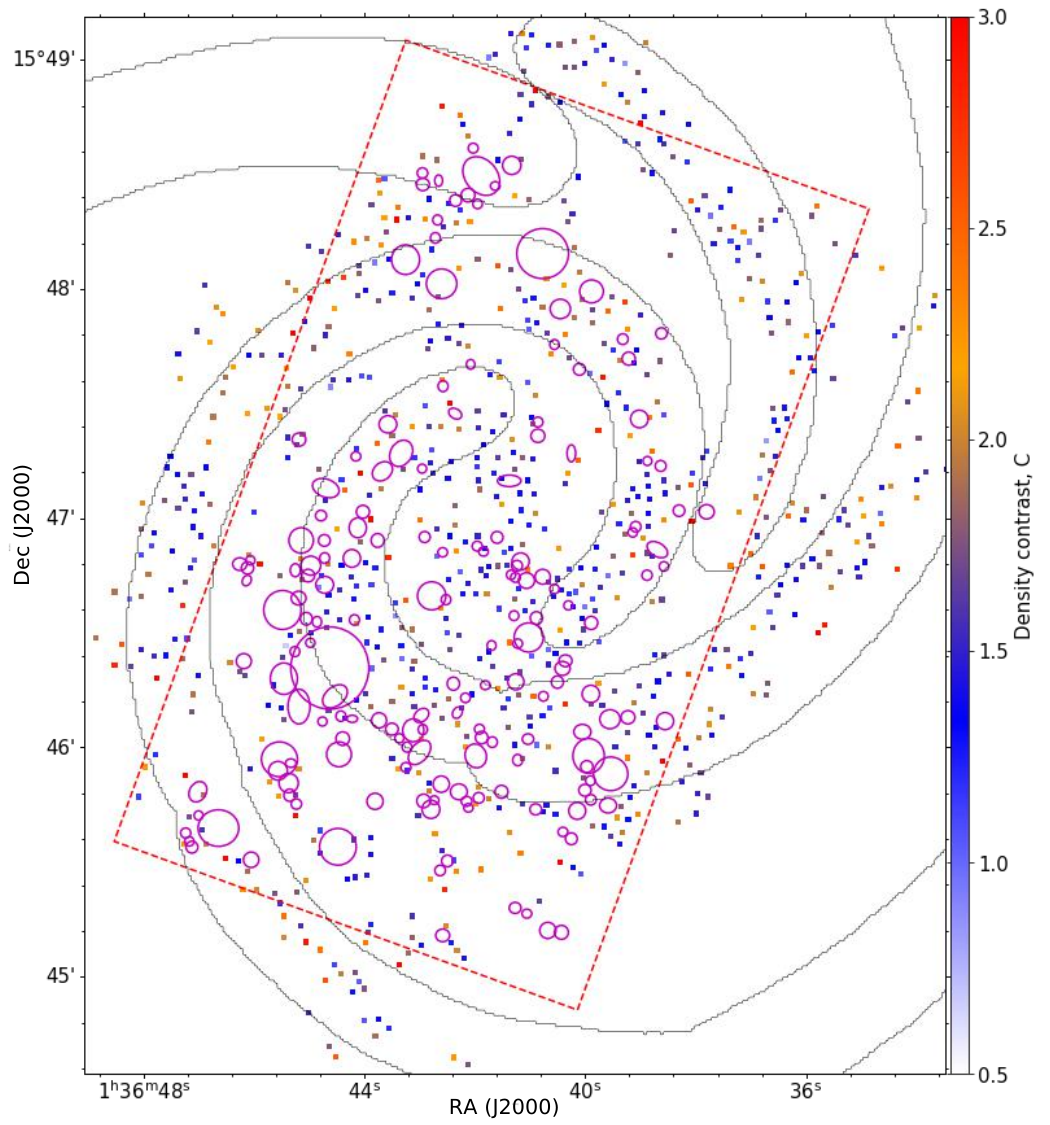}
\caption{The column density contrast map shows the density contrast distribution of all clouds in NGC 628. Each point represents a cloud. Grey contours are the spiral arms identified in \citet{Querejeta2021-656}. Magenta ellipses represent the bubbles identified in \citet{Watkins2023-944}. Red dashed box shows the region observed by JWST.}
\label{contrast}
\end{figure*}

\begin{figure}
\centering
\includegraphics[width = 0.475\textwidth]{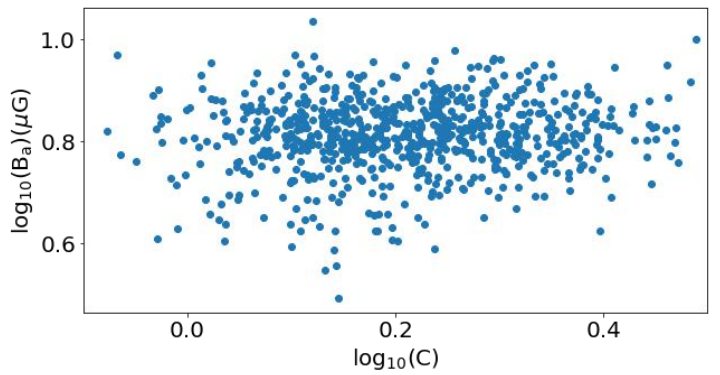}
\caption{Correlation between the average magnetic field strength and the density contrast of the clouds.}
\label{mg}
\end{figure}

We first investigated the impact of spiral arms and bubbles on the density contrast of molecular clouds.
\citet{Querejeta2021-656} identified arms and inter-arms of NGC 628 by using the log-spiral pattern, based on the {\it Spitzer} 3.6 $\mu$m image. The bubbles in NGC 628 were identified in \citet{Watkins2023-944} using JWST mid-infrared observation.
If high-density contrast clouds are triggered by bubbles, bubbles should be larger than the clouds. The median value of the effective radii of all clouds identified in \citet{Zhou2024-534} is $\sim$ 55 pc. Therefore, we only consider the bubbles with the effective radii larger than 55 pc. We construct a density contrast map based on the calculation in \citet{Zhou2024-534}, which shows the density contrast distribution of all clouds in the galaxy. In Fig.\ref{contrast}, each point represents a cloud. We overlay the bubbles and spiral arms on the density contrast map. Overall, high- and low-density contrast clouds are mixed together, primarily concentrated within the spiral arms. Additionally, high-density contrast clouds do not tend to be distributed around bubbles. 
In \citet{Mulcahy2017-600}, the total magnetic field strength of NGC 628 was estimated from the nonthermal emission, assuming equipartition between the energy densities of cosmic rays and the magnetic field, applying the updated formula from \citet{Beck2005-326}. Based on the total magnetic field strength map (Fig.12 in \citet{Mulcahy2017-600}), we roughly estimate the average magnetic field strength for each cloud. As shown in Fig.\ref{mg}, there is no correlation between the magnetic field strength and the density contrast.
Therefore, the spiral arms, bubbles and magnetic fields are not the reasons for the variation in the density contrast of molecular clouds with galactocentric radius.

\subsection{Shear and tidal fields of the galaxy}\label{Sshear}

\begin{figure}
\centering
\includegraphics[width = 0.47\textwidth]{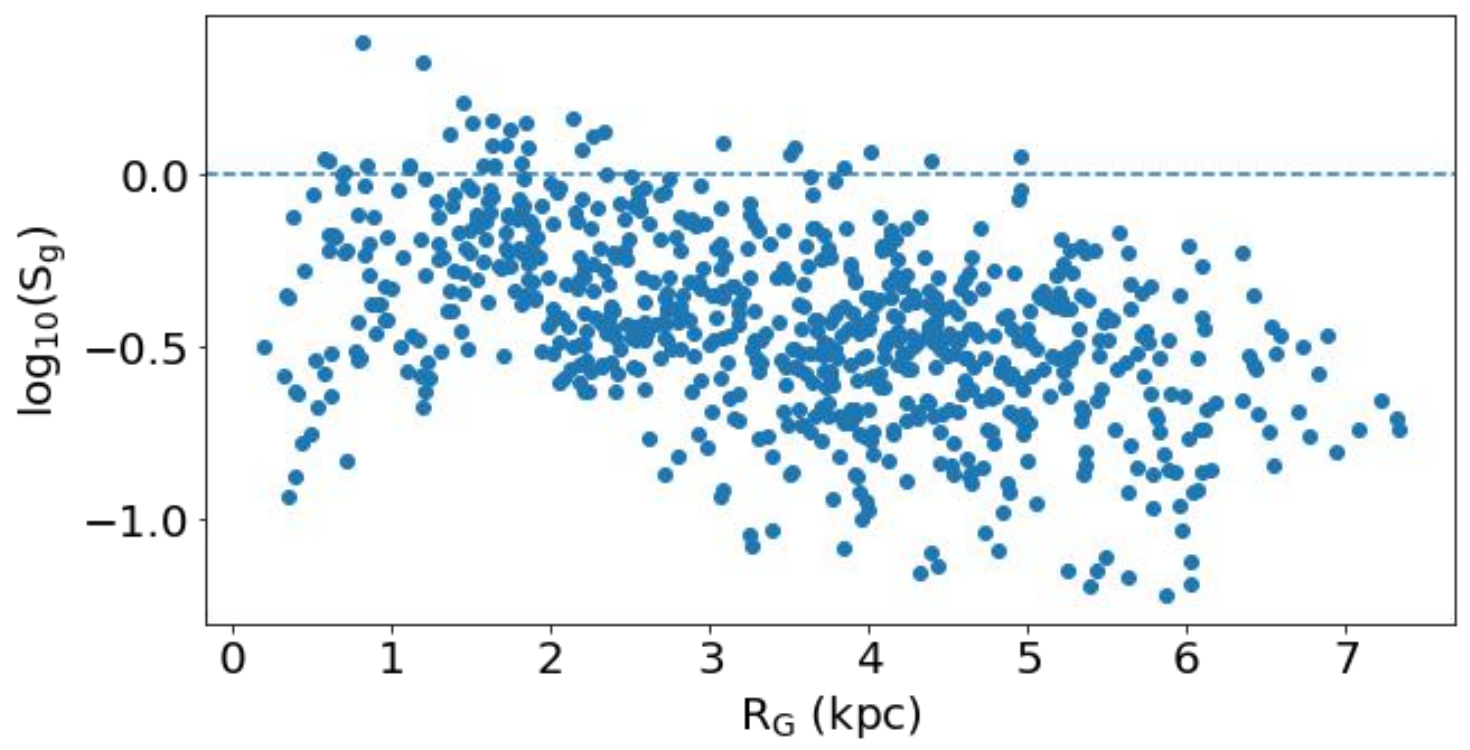}
\caption{Correlation between the shear parameter ($S_{g}$) and the distance to the galaxy center ($R_{\rm G}$).}
\label{shear}
\end{figure}

As discussed in \citet{Hunter1998-493,Dib2012-758,Thilliez2014-31}, 
the shear parameter characterizes the effect of differential rotation on the stability of molecular clouds against gravitational collapse, assessing the time available for perturbations to collapse in the presence of local rotational shear caused by the galaxy's global rotation.
The shear parameter is defined as
\begin{equation}
S_{g} = \frac{\Sigma_{\rm cri}}{\Sigma},
\label{eq:6}
\end{equation}
where $\Sigma$ is the local gas surface density. 
$\Sigma_{\rm cri}$ is the critical surface density, which represents the minimal surface density needed to resist the shear. It is defined as
\begin{equation}
\Sigma_{\rm cri} = \frac{A \sigma \ln(C^{'})}{2\pi G} \, ,
\label{eq:5}
\end{equation}
where $C^{'}$ is taken to be 100 in \citet{Hunter1998-493}.
$\sigma$ is the local gas velocity dispersion. $A$ is the Oort constant, which measures the local shear level. For a spherical cloud of radius $R$, with a rotational velocity $V$ at a radius $R_{G}$, following \citet{Dib2012-758}, $A$ can be calculated as 
\begin{equation}
A =  0.5 \left( \frac{V}{R_{G}} - \frac{dV}{dR_{G}} \right) \, \approx
0.5 \left( \frac{V}{R_{G}} - \frac{\mid V(R_{G}+R) - V(R_{G}-R) \mid}{2R} \right) \, ,
\label{shear}
\end{equation}
where $V$ is determined by the rotation curve of the galaxy. The analytical rotation curve of NGC 628 derived in \citet{Lang2020-897} using CO (2–1) emission is
\begin{equation}
V  =  \frac{2V_{\rm max}}{\pi}\arctan(\frac{R_{G}}{R_{\rm t}}) \approx \frac{2 \times 144.8}{\pi}\arctan(\frac{R_{G}}{0.56~{\rm kpc}})~{\rm km/s},
\label{vrot}
\end{equation}
Shear is effective at tearing apart the gas structure if $S_{g}>1$ and ineffective if $S_{g}<1$.
In Fig.\ref{shear}, only a few structures have $S_{g} >$1, so the effect of shear can be neglected. This is consistent with the results of \citet{Dib2012-758,Thilliez2014-31} for molecular clouds in the Milky Way and the Large Magellanic Cloud. Moreover, there is a clear trend where $S_{g}$ decreases as $R_{G}$ increases, which is inconsistent with the trend in Tab.\ref{prop}, where the proportion of high- and low-density contrast clouds initially increases with $R_{G}$ and then keeps relatively stable.

In a galactic context, the overall gas and stellar distribution can contribute to the confinement or disruption of gas structures. For a cloud at a distance $R_{G}$ from the galactic center, the edge of the cloud from the galactic center is at a distance of $R_{G}$+$R$. 
The tidal acceleration per unit distance (called "tidal strength" in the later) defined in \citet{Stark1978-225} is,
\begin{equation}
T_{G} = \frac{V^{2}}{R_{G}^{2}}-\frac{\partial}{\partial R_{G}}\left(\frac{V^{2}}{R_{G}}\right)  \, .
\label{tg}
\end{equation}
Fig.\ref{tide-co} shows that the tidal strength $T_{G}$ is a strong function of $R_{G}$, which is also inconsistent with the trend in Tab.\ref{prop}.

\subsection{Tidal effects of neighboring material}\label{neighboring}

\subsubsection{Tidal tensor}\label{tensor}

\begin{figure*}
\centering
\includegraphics[width = 0.95\textwidth]{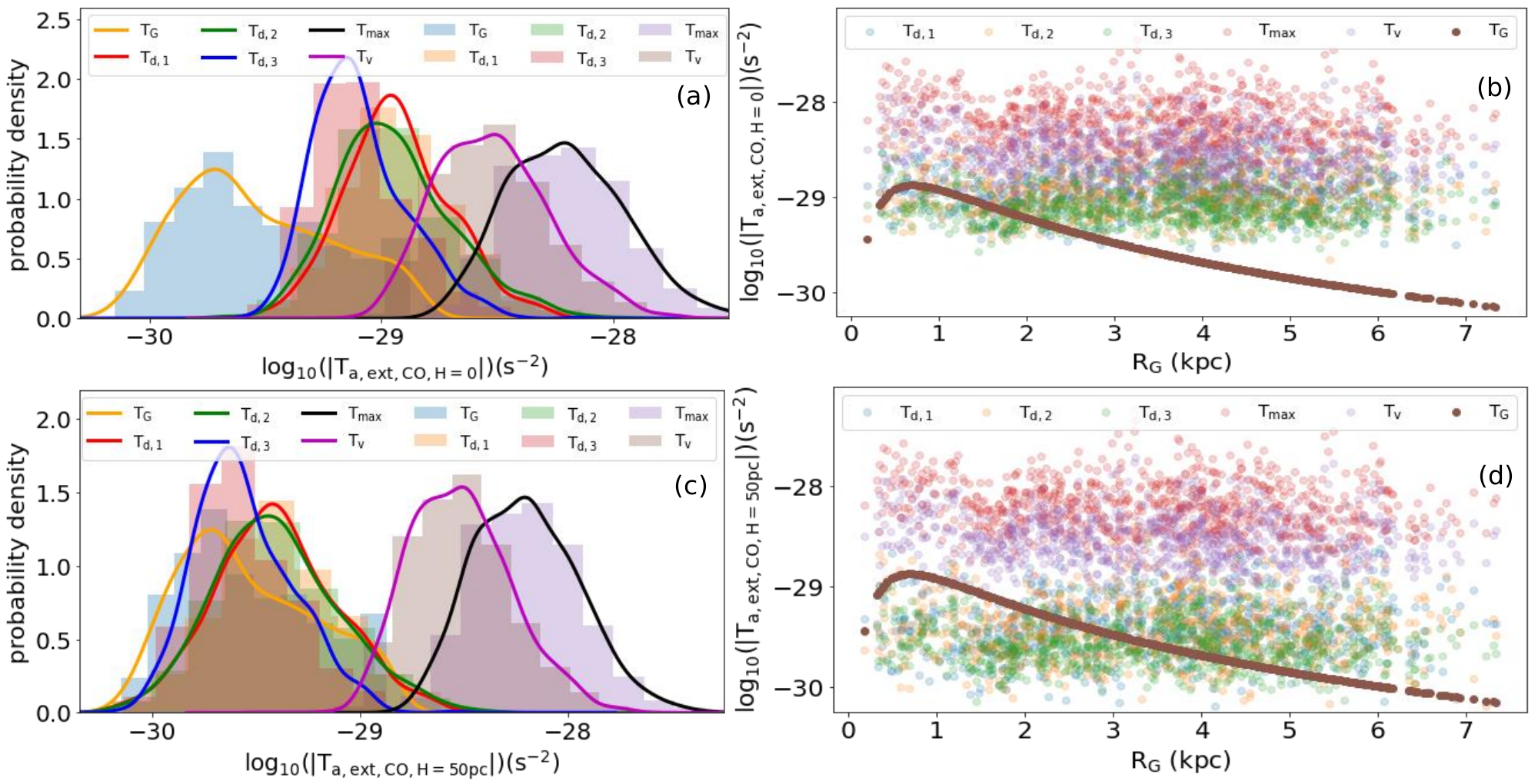}
\caption{Comparison of the tidal strengths obtained from the tidal tensor analysis and the pixel-by-pixel computation based on the CO surface density map. (a) Distribution of the average tidal strength in molecular clouds, assuming the half-thickness of the galaxy $H = 0$. Each type of tidal strength is explained in Tab.\ref{para}; (b) Variation of the average tidal strength of molecular clouds with the distance of the molecular clouds to the galactic center. Panels (c) and (d) are same as panels (a) and (b), but for $H = 50$ pc.}
\label{tide-co}
\end{figure*}

\begin{figure}
\centering
\includegraphics[width = 0.3\textwidth]{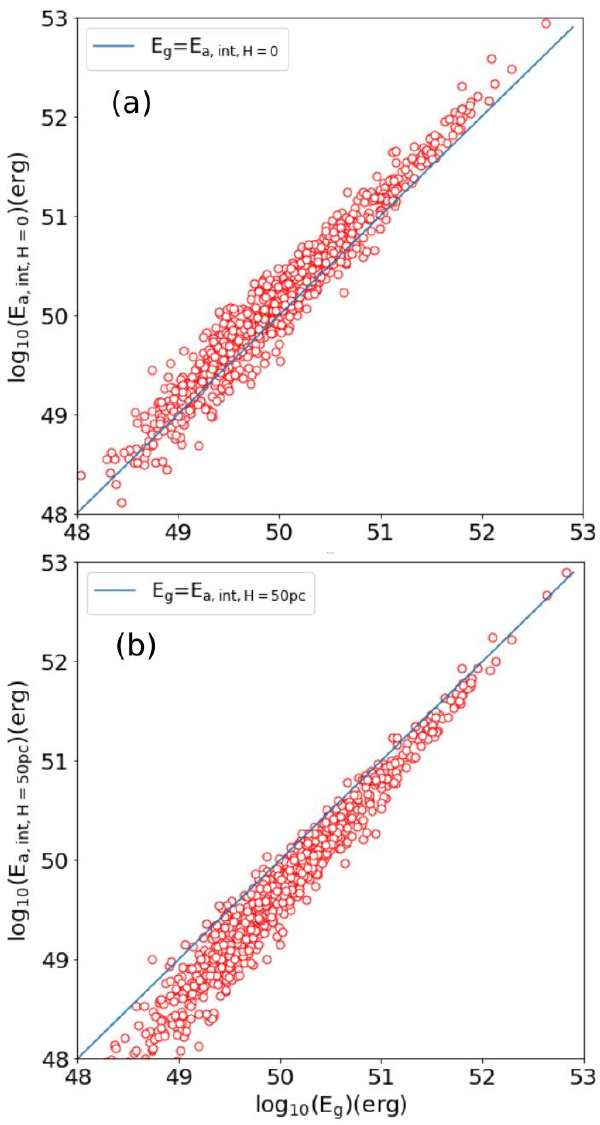}
\caption{Comparison of $E_{\mathrm{a,int,H}}$ and $E_{\mathrm{g}}$ defined in Sec.\ref{tensor}. There are two cases, i.e. $H = 0$ and $H = 50$ pc. These physical quantities are explained in Tab.\ref{para}.}
\label{energy}
\end{figure}

Gravity is a long-range force. A local dense structure evolves under its self-gravity, but as a gravitational center, its gravity can also affect neighboring structures. At the same time, it also experiences the external gravity from neighboring material.
The tidal and gravitational fields are mutually interdependent. 
As described in \citet{Zhou2024PASA,Zhou2024-534}, the hierarchical/multi-scale gravitational coupling of gas structures implies extensive tidal interactions between structures in the galaxy.
Whether the galactic potential has an effect on molecular clouds or not, molecular clouds should be affected by the cumulative tidal interactions of the neighboring material, which may prevent gravitational collapse of the clouds.
\citet{Ramirez2022-515} shows that tidal stresses from neighbouring molecular cloud complex (rather than from the galactic potential) may increase interstellar turbulence.

However, the complex morphology of material distribution in the galaxy means that a complete tidal calculation would be also complex. One should derive the gravitational potential distribution from the observed density distribution and then calculate the tidal field according to the gravitational potential distribution. 
Using the same method as in \citet{Zhou2024-686}, we directly derived the tidal field according to the surface density distribution obtained from CO (2$-$1) emission. Calculating the gravitational potential field $\phi(x, y)$ based on the surface density map $\Sigma(x, y)$ has been discussed in \citet{Gong2011-729,He2023-526}. 
For a layer of half-thickness $H$, the gravitational
potential component $\Phi_{k,\,\mathrm{2D}}$ of surface
density component $\Sigma_{k}$ in Fourier phase space is
\begin{equation}
\Phi_{k,\,\mathrm{2D}}  =  -\frac{2 \pi G \Sigma_{k}}
{\left| {k}\right|(1 + \left|{k H}\right|)},
\label{2D}
\end{equation}
where $\left|k\right| = \sqrt{k_x^2+k_y^2}$.  Note that
for $|k H| \gg 1$, $\Phi_{k,\,\mathrm{2D}}
\sim -4 \pi G \rho_k /k^2$, which is the solution of the Poisson equation in three dimensions, for $\rho_k  = 
\Sigma_k/2H$. When $|k H| \ll 1$,
Eqn.\ref{2D} is the solution of the Poisson equation for an infinitesimally thin layer. The gravitational potential
$\Phi_\mathrm{2D}(x,y)$ is the inverse Fourier transform of $\Phi_{k,\,\mathrm{2D}}$.
The tidal tensor $\mathbf{T}$, defined as $T_{ij}  =  \frac{\partial \Phi}{\partial i \partial j}$, only has two eigenvalues in the case of 2D, i.e. $\lambda_{i}$ ($i$ = 1,2). 
The tidal tensor of a structure includes both self-gravity and external tides, that is, $\mathbf{T} =  \mathbf{T}_{\mathrm{ext}} + \mathbf{T}_{\mathrm{int}}$. 
The decomposition of the tidal tensor is discussed in detail in \citet{Zhou2024-686}.
For three extreme cases, the external tides can be estimated by $\lambda_{1}$, $\lambda_{2}$, and $(\lambda_{2}-\lambda_{1})/2$, and they are called $T_{\rm d,1}$, $T_{\rm d,2}$, and $T_{\rm d,3}$, respectively. 

For the surface density map, the half-thickness $H$ is 0. Taking a characteristic scale height of the molecular gas in the galaxy as 50 pc
\citep{Bronfman1988-324,Stark2005-619,Cox2005-43,Ellsworth2013-770,Jeffreson2022-515}, which is also close to the median value of the effective radii of all identified clouds ($\sim$55 pc), then we have $H = 50$ pc. We calculated the tidal tensors for both $H = 0$ and $H = 50$ pc. Fig.\ref{tensor-co} displays different components of the tidal tensor in each case. 

In the case of 3D, for a cloud, the energy derived from $\mathbf{T}_{\mathrm{int}}$ should be comparable with its self-gravity energy ($E_{\rm g}$), i.e. Eqn.(A8) in \citet{Zhou2024-686},
\begin{equation}
E_{\rm a,int,H} \approx \frac{1}{2}
M *\mathrm{Tr(\mathbf{T}_{a,int,H})}* R^{2} \approx \frac{GM^{2}}{R} \approx E_{\rm g}.
\label{Et}
\end{equation}
In Fig.\ref{energy}, $E_{\rm a,int,H = 0}$ is comparable with $E_{\rm g}$. However, $E_{\rm a,int,H = 50pc}$ is systematically smaller than $E_{\rm g}$. Thus, the computed 2D tidal field strength ($H = 0$) can roughly reflect the real 3D tidal field strength of the galaxy.


\subsubsection{Pixel-by-pixel computation}\label{pp}

\begin{figure}
\centering
\includegraphics[width = 0.48\textwidth]{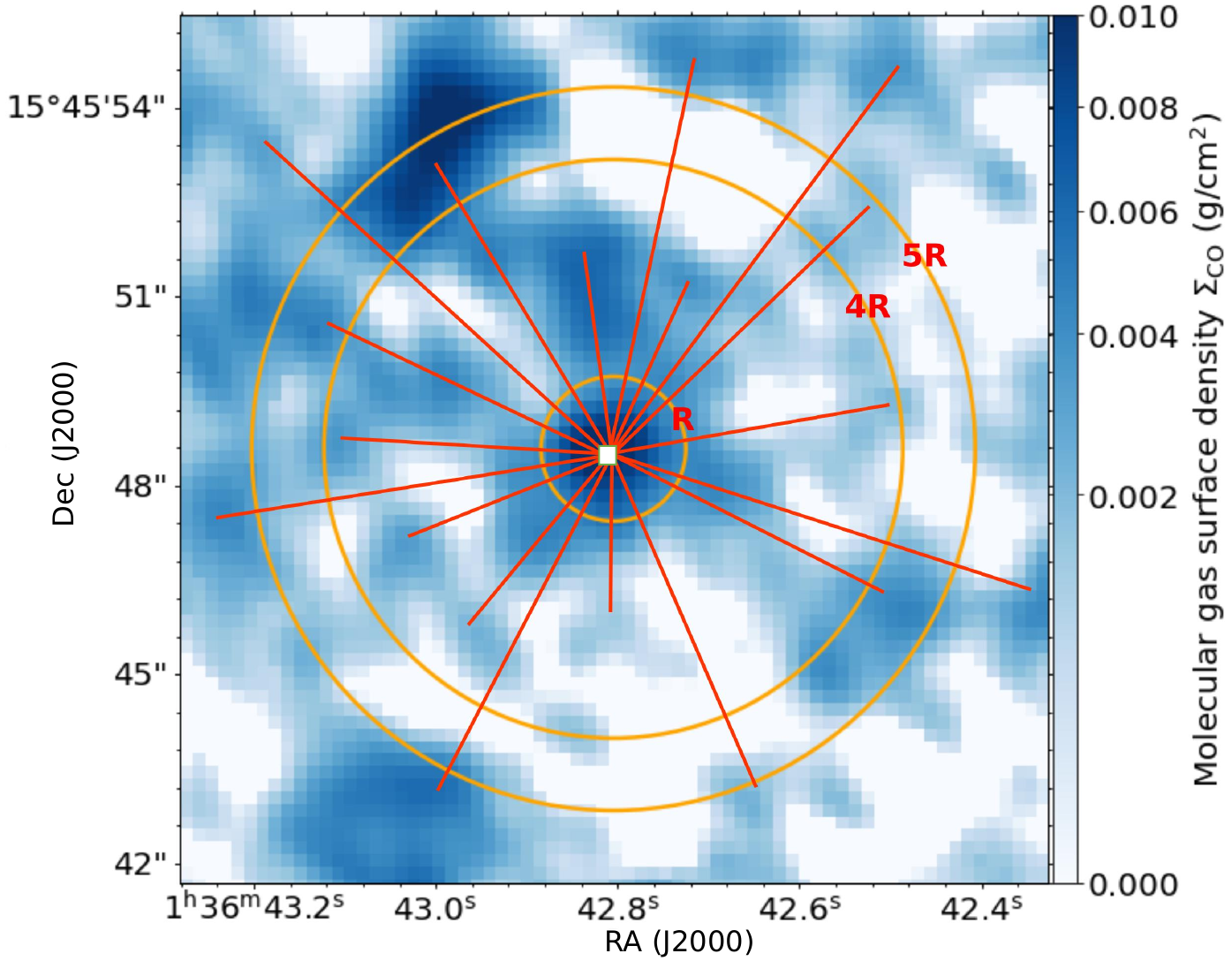}
\caption{Illustration of the tidal effects exerted by the neighboring material on a pixel/point within a cloud. $R$ is the effective radius of the cloud.}
\label{case}
\end{figure}

\begin{figure*}
\centering
\includegraphics[width = 0.95\textwidth]{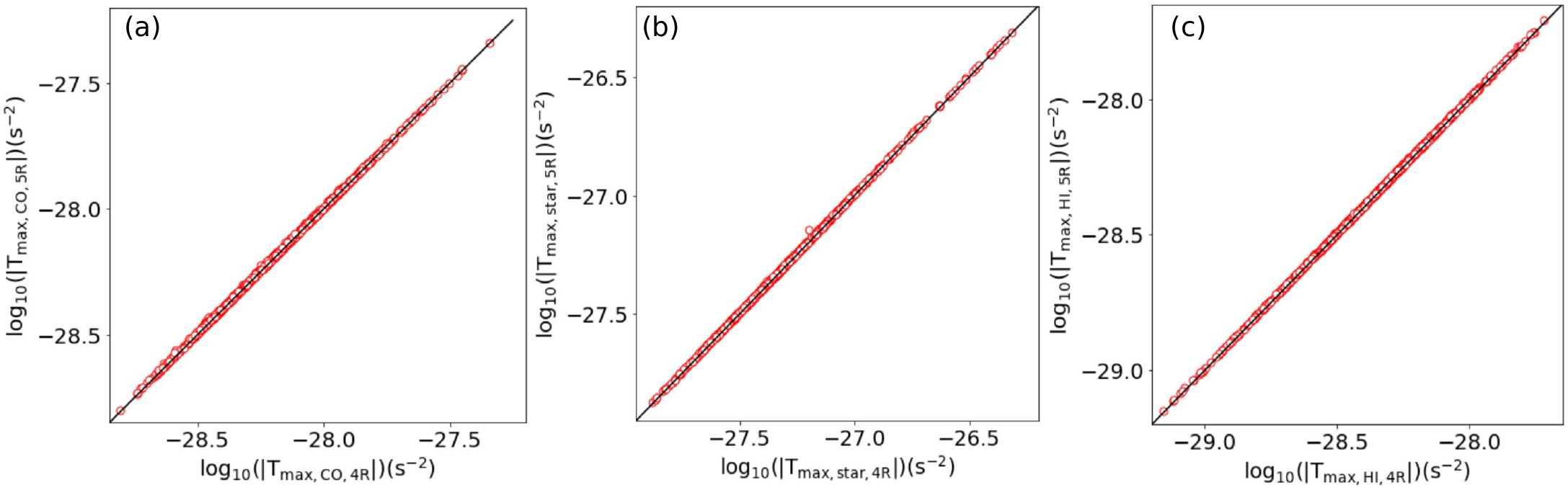}
\caption{Comparison of the tidal strength exerted on molecular clouds by neighboring material within 4 times and 5 times the effective radius of the clouds.}
\label{45}
\end{figure*}

In this section, we further verify the above tidal tensor analysis through an original methodology. Tidal forces can significantly contribute to the deformation and intricate shapes of the gas structures. At a distance $R^{'}$ from the center of cloud B with mass $M^{'}$, the tidal strength sustained by cloud A due to the external gravity of cloud B is
\begin{equation}
T \approx \frac{2GM^{'}}{R^{'3}}.
\label{point}
\end{equation}
The cumulative tidal strength at a pixel signifies the aggregate deformation resulting from external gravity at that point. Unlike a rigid body or a point mass, a gas structure is flexible and can deform. Gas structures are often irregular in shape, and their morphologies can be quite intricate. While the gravitational forces acting on a gas structure from different directions might balance out, the tidal effects within the structure, driven by external gravity, do not. To quantify the collective tidal strength within a structure induced by external gravity of the neighboring material, we adopt the scalar superposition.

We divided the surface density map of the galaxy into two distinct regions using the cloud's mask obtained in \citet{Zhou2024-534}, i.e. the cloud itself and all other material outside of it. To compute the external tidal effects exerted by the surrounding material on a point within the cloud, we use Eqn.\ref{point} to directly calculate the tidal strength at that point pixel by pixel, as depicted in Fig.\ref{case}. The average tidal strength across all pixels within the cloud subsequently provides an estimate of the average tidal strength experienced by the cloud. Our computation is limited to a 2D plane. Actually, there is a characteristic scale height of the molecular gas in the galaxy. In Eqn.\ref{point}, the tidal strength is very sensitive to the distance. If the external material and the calculated point in a cloud do not lie on the same plane, the 2D computation results in an underestimation of the distance between them, consequently leading to an overestimation of the tidal strength (the maximum computation, $T_{\rm max}$). 

A mask of a cloud contains $N$ pixels. 
Since the tidal strength $T$ is very sensitive to the distance $R^{'}$, material located sufficiently far from a cloud can be expected to have a negligible tidal effects on the cloud. Therefore, it is not necessary to account for all external material within the entire galaxy. We only need to consider the material near the cloud. As illustrated in Fig.\ref{case},
we first calculated the average tidal strength exerted by the material within a radius of 4 times the effective radius of a cloud, with the center of the cloud as the origin. Then, we extended the calculation to a radius of 5 times the effective radius of the cloud and performed the same calculation. As shown in Fig.\ref{45}, for all clouds in the galaxy, both calculations yield similar results, indicating that the tidal effects of material beyond 4 times the effective radius of a cloud can be neglected.

To compare with the scalar superposition, we also calculate a vector superposition of tides, enabling direct cancellation of tidal interactions. By establishing a 2D coordinate system with the calculated pixel/point in a cloud as the origin, we decomposed all external tides along the coordinate axes. Subsequently, we calculated the aggregate tidal strength at the calculated pixel after performing vector superposition along each axis ($T_{\rm v}$).

\subsubsection{Baryonic matter}

\begin{figure*}
\centering
\includegraphics[width = 0.95\textwidth]{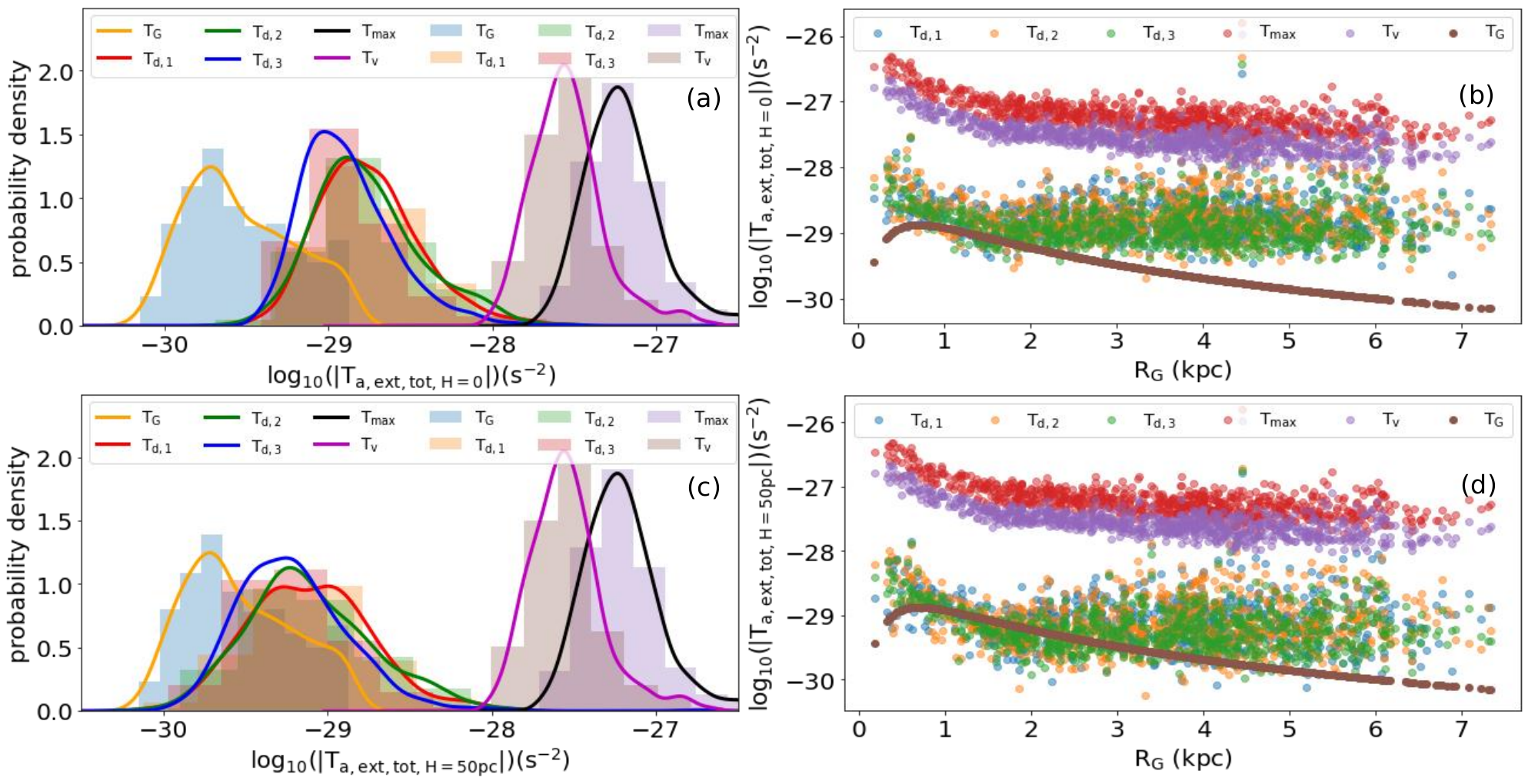}
\caption{Same as Fig.\ref{tide-co}, but all tidal strengths derived from the total surface density map, rather than only the CO surface density map in Fig.\ref{tide-co}.}
\label{tide-tot}
\end{figure*}

Although we focus solely on molecular clouds as sites of star formation in the galaxy, it's inadequate to only consider molecular gas when accounting for external tidal effects on the clouds. 
Strictly speaking, we should include all baryonic matter, and even dark matter. 
As a rough approximation, apart from molecular gas surface density ($\Sigma_{\mathrm{CO}}$), we also consider atomic gas surface density ($\Sigma_{\mathrm{HI}}$) and stellar mass surface density ($\Sigma_{\mathrm{*}}$). 

Using the recipes in \citet{Leroy2021-257,Sun2022-164},
assuming optically thin 21~cm emission, the 21~cm line integrated intensity $I_{\rm HI}$ can be converted to atomic gas surface density $\Sigma_{\rm HI}$ via
\begin{equation}
    \frac{\Sigma_{\rm HI}}{\rm M_{\odot}~pc^{-2}}  =  2.0\times10^{-2} \left(\frac{I_{\rm HI}}{\rm K~km s^{-1}}\right).
    \label{hi}
\end{equation}
Near-IR imaging data can be used to trace the old stellar mass distribution. Using the \textit{Spitzer} IRAC $3.6$~$\mu$m image, the stellar continuum intensity at $3.6$~$\mu$m can be converted to the stellar mass surface density by
\begin{equation}
    \frac{\Sigma_{\rm *}}{\rm M_{\odot}~pc^{-2}}  =  350 \left(\frac{\gamma_{\rm 3.4\mu m}}{0.5}\right)
    \left(\frac{I_{\rm 3.6\mu m}}{\rm MJy~sr^{-1}}\right).
    \label{hi}
\end{equation}
Here, $\gamma_{\rm 3.4\mu m}$ is the stellar mass-to-light ratio at $3.4$~$\mu$m, we take a value of 0.5 $M_{\odot}/L_{\odot}$ \citep{Leroy2021-257}.
As shown in Fig.\ref{surface}, the stellar mass surface density is significantly higher than molecular and atomic gas surface density. We combined $\Sigma_{\mathrm{CO}}$, $\Sigma_{\mathrm{HI}}$ and $\Sigma_{\mathrm{*}}$ to create a total surface density map to approximate the baryonic matter distribution in the galaxy, which is dominated by $\Sigma_{\mathrm{*}}$. 


\subsection{Comparison of different tidal calculations}
\begin{table*}
\centering
\caption{Explanation of the physical quantities. The tidal strength is the tidal acceleration per unit distance. In the tidal tensor analysis, the tidal strength is the numerical value of a specific component on the tidal tensor.}
\begin{tabular}{cc}
\hline
$R_{G}$ & The distance to the galactic center.\\
$\mathrm{Tr(\mathbf{T}_{int})}$ & The trace of 2D tidal tensor derived from the CO surface density in Sec.\ref{tensor}. \\
$E_{\mathrm{a,int,H}}$ & The energy corresponding to $\mathrm{Tr(\mathbf{T}_{a,int,H})}$ averaged over a cloud. $H$ is the half-thickness of the galaxy. \\ &There are two cases, i.e. $H = 0$ and $H =  50$ pc. \\
$E_{\mathrm{g}}$ & Self-gravity energy of a cloud. \\
$T_{G}$ & The tidal strength calculated using the rotation curve by Eqn.\ref{tg}. \\
$T_{\rm d,1}$, $T_{\rm d,2}$ and $T_{\rm d,3}$ & Three extreme cases of the tidal strengths obtained from the decomposition of the tidal tensor. \\
$T_{\rm max}$ and $T_{\rm v}$ & The tidal strengths computed pixel-by-pixel on 2D plane in the cases of scalar and vector superpositions, respectively, \\
&see Sec.\ref{pp} for more details.\\

\hline
\label{para}
\end{tabular}
\end{table*}

Some physical quantities calculated in Sec.\ref{neighboring} are summarized in Tab.\ref{para}. Here, we will compare the different types of tidal calculations mentioned above.

\subsubsection{Only the molecular gas}
For the tidal strengths calculated according to the CO surface density map, Fig.\ref{tide-co}(a) shows that the components $T_{\rm d,1}$, $T_{\rm d,2}$, and $T_{\rm d,3}$ derived from the tidal tensor analysis in Sec.\ref{tensor} are comparable.
As expected, the values of $T_{\rm max}$ are significantly larger than others. The tidal strength $T_{G}$ derived in Sec.\ref{Sshear} according to the rotation curve is significantly smaller than others at large $R_{G}$, when $H = 0$.
In Fig.\ref{tide-co}(b), we plot the variation of the tidal strengths with $R_{G}$. Except for $T_{G}$, the other quantities do not vary with the distance to the galaxy center. Because the calculations of the other quantities are based on the CO surface density map. In Fig.\ref{surface}(a), the galactic center does not have significantly higher CO surface density. 

\subsubsection{All matter}
For the tidal strengths calculated based on the total surface density map, although the components $T_{\rm d,1,tot}$, $T_{\rm d,2,tot}$, and $T_{\rm d,3,tot}$ derived from the tidal tensor analysis in Sec.\ref{tensor} are comparable, they are significantly smaller than the tidal strengths obtained from pixel-by-pixel computation ($T_{\rm v,tot}$ and $T_{\rm max,tot}$) in Sec.\ref{pp}. In Fig.\ref{tide-tot}(b), all calculated tidal strengths change with $R_{G}$, and the scatterings are also smaller than those in Fig.\ref{tide-co}(b). Especially, in Fig.\ref{tide-tot}(b), when $R_{\rm G} <$ 4 kpc, $T_{\rm v,tot}$ and $T_{\rm max,tot}$ decrease as $R_{\rm G}$ increases, and then remains relatively constant when $R_{\rm G} >$ 4 kpc. This aligns well with the dependence of the proportion of high- and low-density contrast clouds on $R_{\rm G}$, as presented in Sec.\ref{C}. 
\subsubsection{Scale height and scalar versus vector superpositions}

In the two cases mentioned above, $H = 50$ pc generates lower tidal strengths than $H = 0$, but is otherwise identical to $H = 0$ in all other respects.
As described in Sec.\ref{pp}, $T_{\rm v,tot}$ and $T_{\rm max, tot}$ are calculated in ways that are quite different, but their trends of variation with respect to $R_{\rm G}$ are consistent. The only difference is in the magnitude of the tidal strengths.

\subsubsection{Tidal tensor analysis versus pixel-by-pixel computation}
The pixel-by-pixel computation in Sec.\ref{pp} can reflect the variation in tidal strength with $R_{\rm G}$ more precisely than the tidal tensor analysis in Sec.\ref{tensor}, as it has a smaller scatter in tidal strength, as shown in Fig.\ref{tide-tot}.
However, to accurately estimate the tidal strength using the pixel-by-pixel computation, we need to know the specific 3D spatial distribution of the material in the galaxy, which is challenging. 

\subsubsection{Consequences of the tidal analysis}
Above, we conducted an extensive discussion aimed at summarizing an effective tidal analysis method for future research. To explain the variation in the density contrast of molecular clouds with $R_{\rm G}$ in NGC 628, it is necessary to integrate a total (baryonic matter) surface density map that includes molecular gas, atomic gas, and stellar components. This is because the stellar mass surface density is significantly higher than the molecular gas surface density, indicating stronger tidal effects. The pixel-by-pixel computation can more precisely reflect the variation in tidal strength with $R_{\rm G}$ than the tidal tensor analysis but cannot provide a correct magnitude estimate of the tidal strength due to the lack of knowledge about the specific 3D spatial distribution of the material in the galaxy. This issue can be resolved using the tidal tensor analysis. Therefore, by combining these two methods, we can achieve both appropriate
 variation and magnitude estimates of the tidal strength across the galaxy.

\section{Discussion and conclusion}

The physical factors influencing the density contrast of molecular clouds are the same ones that impact star formation in the galaxy. The proportion of high- and low-density contrast clouds initially increases with the distance to the galactic center ($R_{G}$) and then keeps relatively stable. The spiral arms, bubbles and magnetic fields are not responsible for the variations in density contrast observed among molecular clouds. The galactic shear and tidal strength calculated by the rotation curve consistently decrease as $R_{G}$ increases, inconsistent with the variation of the proportion between high- and low-density contrast clouds with $R_{G}$. 

We studied the tidal effects of neighboring material on each cloud using two different methods, i.e. the tidal tensor analysis and the pixel-by-pixel computation. All calculations were carried out on 2D plane, but the computed 2D tidal field strength ($H = 0$) in the tidal tensor analysis can roughly reflect the real 3D tidal field strength of the galaxy. In the pixel-by-pixel computation, we consider both scalar and vector superpositions of tides. The tidal effects of material beyond 4 times the effective radius of a cloud can be neglected. Although we focus solely on molecular clouds as sites of star formation in the galaxy, it's inadequate to only consider molecular gas when accounting for external tidal effects on the clouds. We constructed a total surface density map by integrating molecular gas, atomic gas, and stellar mass surface density maps to estimate the distribution of baryonic matter within the galaxy.

In all cases, assuming a galaxy half-thickness of $H = 50$ pc results in lower tidal strengths compared to $H = 0$, while remaining otherwise identical to $H = 0$ in every other aspect. In both tidal tensor analysis and pixel-by-pixel computation, the tidal strengths calculated according to the CO (2-1) surface density map do not vary with $R_{G}$, because the galactic center does not have significantly higher CO surface density. However, for the tidal strengths calculated based on the total surface density map, all of them change with $R_{G}$. The tidal strengths derived from the pixel-by-pixel computation have smaller scatter than those obtained from tidal tensor analysis and can reflect the variation in tidal strength with $R_{\rm G}$ more precisely. When $R_{\rm G} <$ 4 kpc, the tidal strengths calculated pixel-by-pixel decrease as $R_{\rm G}$ increases, and then stabilize for $R_{\rm G} >$ 4 kpc. This trend aligns closely with the variation in the proportion of high- and low-density contrast clouds with $R_{\rm G}$. Therefore, the tidal effects of neighboring material have a significant impact on the development of molecular cloud's density contrast, which in turn affects star formation within molecular cloud.

In Fig.\ref{tide-tot}, $T_{G}$ calculated by the rotation curve consistently decreases with increasing $R_{\rm G}$. $T_{\rm v,tot}$ and $T_{\rm max,tot}$ represent local tidal strengths from neighboring material and are only sensitive to local material distribution. As shown in Fig.\ref{surface}, the surface density is significantly higher in galactic center. Therefore, $T_{\rm v,tot}$ and $T_{\rm max,tot}$ only decrease with increasing $R_{\rm G}$ in the central region of the galaxy. And they remains relatively constant in the outer regions of the galaxy, as the surface density distribution in the outer regions is relatively uniform. Therefore, one important reason for the low star formation rate in the galactic center is that the tidal effects from neighboring material acting on molecular clouds are strong, restricting the gravitational collapse of the clouds. As a result, the clouds exhibit low density contrasts.

Molecular clouds do not evolve in isolation, their gravitational collapse is modulated by various external physical processes. Although the measured velocity gradients in \citet{Zhou2024PASA,Zhou2024-534} support the gravitational collapse of gas structures on galaxy-cloud scales, the collapse is significantly slower than a pure free-fall gravitational collapse. In the case of free-fall, the velocity gradient ($\nabla v$) and the scale ($R$) satisfy $\nabla v \propto R^{-1.5}$. However, for NGC 628, the fitting in \citet{Zhou2024-534} was $\nabla v \propto R^{-0.9}$. For NGC 4321 and NGC 5236, \citet{Zhou2024PASA} obtained a similar slope, i.e. $\nabla v \propto R^{-0.8}$. As discussed in \citet{Zhou2024PASA}, the slowing down of a pure free-fall gravitational collapse can be caused by multiple physical processes. The results of this work suggest that the tidal effects of neighboring material can resist the gravitational collapse of molecular clouds.

This work is merely a case study, and future work needs to apply similar methods to large galaxy samples containing different types of galaxies to verify the generality of the conclusions presented here.

\section{Data availability}

All the data used in this work are available on the PHANGS team website.
\footnote{\url{https://sites.google.com/view/phangs/home}}.

\section*{Acknowledgements}
Thanks the referee for providing detailed review comments, which have helped to improve and clarify this work.
We would like to thank R. Beck for sharing the nonthermal emission map of NGC 628 with us.
It is a pleasure to thank the PHANGS team, the data cubes and other data products shared by the team make this work can be carried out easily. This paper makes use of the following ALMA data: ADS/JAO.ALMA\#2012.1.00650.S and ADS/JAO.ALMA\#2017.1.00886.L.
ALMA is a partnership of ESO (representing its member states), NSF (USA) and NINS (Japan), together with NRC (Canada), NSTC and ASIAA (Taiwan), and KASI (Republic of Korea), in cooperation with the Republic of Chile.

\bibliography{ref}
\bibliographystyle{aasjournal}

\begin{appendix}
\twocolumn

\section{Supplementary maps}\label{map}

\begin{figure*}
\centering
\includegraphics[width = 0.95\textwidth]{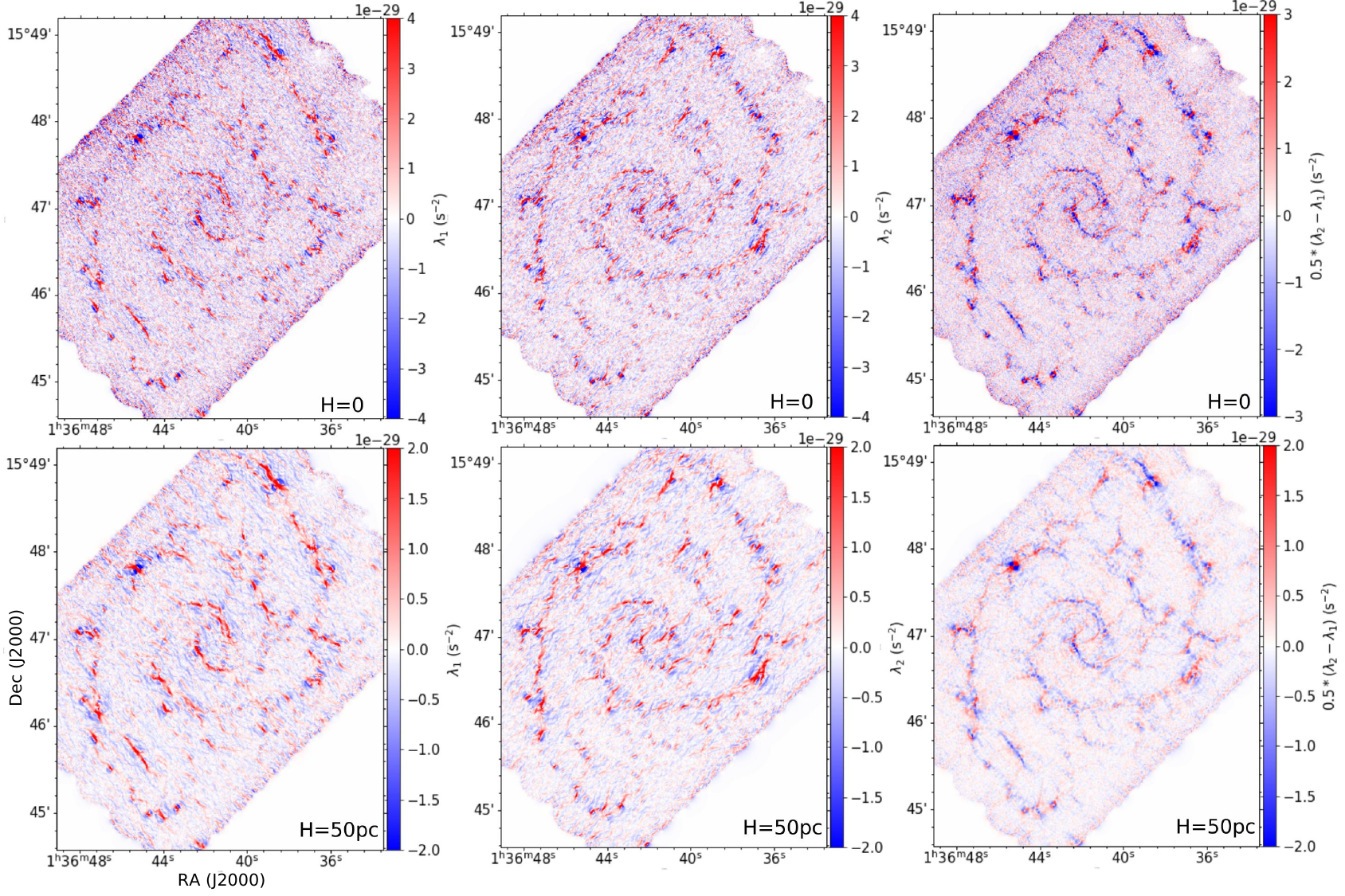}
\caption{Different components of the tidal tensor derived from the CO surface density map. They are three extreme cases of the external tides, as described in Sec.\ref{tensor}.}
\label{tensor-co}
\end{figure*}
\begin{figure*}
\centering
\includegraphics[width = 0.95\textwidth]{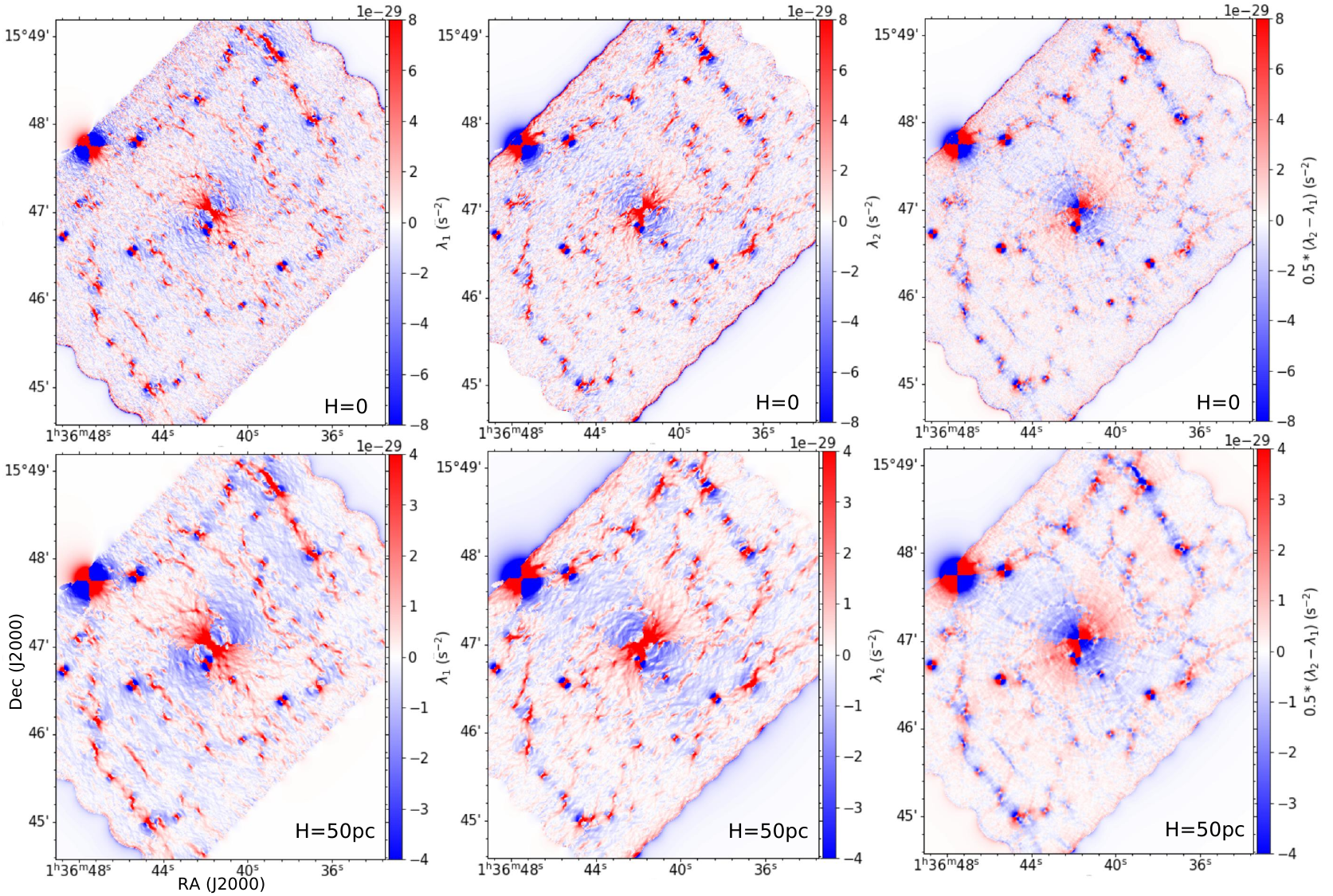}
\caption{Same as Fig.\ref{tensor-co}, but derived from the total surface density map.}
\label{tensor-tot}
\end{figure*}

\end{appendix}

\clearpage
\noindent
\end{document}